\documentclass[a4paper,11pt]{article}
\usepackage{pos,epsfig}
\usepackage{fancyvrb} % for "\Verb" macro
\VerbatimFootnotes    % enable use of \Verb in footnotes
\usepackage{xcolor}
\usepackage{cancel}

\def \mb      {\texttt{MB{}}{}}
\def \pltest      {\texttt{PlanarityTest}{}}

\def \mbm  {\texttt{MB.m}{}}
\def \mbr  {\texttt{MB}{}}

\def\mathswitch#1{\relax\ifmmode#1\else$#1$\fi}
\def\mathswitchr#1{\relax\ifmmode{\mathrm{#1}}\else$\mathrm{#1}$\fi}

\newcommand{\as}{\alpha_{\mathrm s}}

\newcommand{\seff}[1]{\sin^2\theta_{\rm eff}^{#1}}
\newcommand{\PW}{\mathswitchr W}
\newcommand{\PZ}{\mathswitchr Z}

\newcommand{\Pt}{\mathswitchr t}

\newcommand{\MW}{\mathswitch {M_\PW}}

\newcommand{\GZ}{\mathswitch {\Gamma_\PZ}}

\newcommand{\OO}{{\mathcal O}}

\title{%
{\flushright{ 
\small \texttt{DESY 20-184}
\\
\small \texttt{KW 20-002}
\\[8mm]
}}
Electroweak precision pseudo-observables
at 
the $e^+e^-$ Z-resonance 
peak 
}
\ShortTitle{Electroweak precision pseudo-observables at the $e^+e^-$ Z-resonance peak}

\author[a]{Ievgen Dubovyk}
\author[b]{Ayres Freitas}
\author*[a]{Janusz Gluza}
\author[a]{Krzysztof Grzanka}
\author[a,d]{\mbox{Tord Riemann}} 
\author[c]{Johann Usovitsch}

\affiliation[a]{Institute of Physics, University of Silesia, 
Katowice, Poland}

\affiliation[b]{Pittsburgh Particle physics, Astrophysics \& Cosmology Center
(PITT PACC),\\ Department of Physics \& Astronomy, University of Pittsburgh, Pittsburgh, PA 15260, USA}

\affiliation[c]{PRISMA Cluster of Excellence, Institut f\"ur Physik, \\Johannes Gutenberg-Universit\"at Mainz,
D-55099 Mainz, Germany
}

\affiliation[d]{%
DESY, 15738 Zeuthen, Germany
}

\emailAdd{ievgen.dubovyk@us.edu.pl}
\emailAdd{afreitas@pitt.edu}
\emailAdd{janusz.gluza@us.edu.pl}
\emailAdd{krzysztof.grzanka@us.edu.pl}
\emailAdd{tordriemann@gmail.com}
\emailAdd{jusovitsch@googlemail.com}

\abstract{Phenomenologically relevant electroweak precision pseudo-observables related to Z-boson physics are discussed in the context of the strong experimental demands of future $e^+e^-$ colliders.
The recent completion of two-loop Z-boson results is summarized and a prospect for the 3-loop Standard Model calculation of the Z-boson decay pseudo-observable is given.
}

\FullConference{%
  40th International Conference on High Energy physics - ICHEP2020\\
  July 28 - August 6, 2020\\
  Prague, Czech Republic (virtual meeting)
}

\begin{document}
\maketitle

%\section{$Z-$boson physics: Past, present, future}

One of the exciting activities in searching for non-standard effects in particle physics is the precision study of the $Z-$-boson decay in $e^+e^-$ collisions.  Electron-positron collisions form the $Z$ resonance at center-of-mass energies around $91$ GeV.  This process was instrumental in the \texttt{LEP} era, leading to the detailed knowledge of crucial parts of the Standard Model (SM) \cite{ALEPH:2005ab,Bardin:1995-YR03}.
Up to $5 \times 10^{12}$ $Z$-boson decays are planned to be 
observed at the $Z$-boson resonance with the  FCC-ee collider \cite{Abada:2019zxq,Blondel:2019qlh}, while it would be about one order of magnitude less at the CEPC \cite{CEPC-SPPCStudyGroup:2015csa}. 
These statistics are about six orders of magnitude larger than at LEP and may lead to very accurate experimental measurements of the so-called Electro-Weak Pseudo-Observables (EWPOs), if the systematic experimental errors can be hold appropriately small. In turn, this means that theoretical predictions must also be very exact, of the order of 3- to 4-loop QCD and EW effects \cite{Blondel:2018mad}. This level of accuracy and potential distortions from the SM predictions will put stringent limits on theory scenarios beyond the SM with New Physics virtual particles and interactions.
A substantial step in this direction of accuracy within the SM was a recent  calculation of the most difficult massive bosonic two-loop contributions to the $Z$-boson decay \cite{Dubovyk:2016aqv,Dubovyk:2018rlg,Dubovyk:2019szj}. In this way, the Standard Model electroweak two-loop corrections are completed. The focus can be directed now on the next, NNNLO order of loop calculations. 
Their contributions will be necessary in order to meet the anticipated experimental accuracies.

%\section{Completion of the $Z-$boson decay SM EWPO calculations at the 2-loop level.}

\begin{center}
%-------------------------------------------------------------
{\footnotesize{ 
\begin{table*}[b] %[h!]  %tab.2 PLB2018
\renewcommand{\arraystretch}{1.2}
\begin{center}
\begin{tabular}{|l|r|r|r|r|r|r|}
\hline
\multicolumn{1}{|r|}{$\Gamma_i$ [MeV]} & $\Gamma_e\;\;$ & $\Gamma_\nu\;\;$ & $\Gamma_d\;\;$ & $\Gamma_u\;\;$ & 
 $\Gamma_b\;\;$ & $\Gamma_\PZ\;\;$ \\
\hline \hline
Born & 81.142 & 160.096 & 371.141 & 292.445 & 369.562 & 2420.19
\\
$\OO(\alpha)$ & 2.273 & 6.174 & 9.717 & 5.799 & 3.857 & 60.22 
\\
$\OO(\alpha\as)$ & 0.288 & 0.458 & 1.276 & 1.156 & 2.006 & 9.11 
\\
$\OO(\alpha_\Pt\as^2,\,\alpha_\Pt\as^3,\,\alpha^2_\Pt\as,\,\alpha_\Pt^3)$ &
 0.038 & 0.059 & 0.191 & 0.170 & 0.190 & 1.20 \\
$\OO(N_f^2\alpha^2)$ & 0.244 & 0.416 & 0.698 & 0.528 & 0.694 & 5.13 
\\
$\OO(N_f\alpha^2)$ & 0.120 & 0.185 & 0.493 & 0.494 & 0.144 & 3.04 \\
$\OO(\alpha^2_{\rm bos})$ & 
                    0.017 & 0.019 & 0.059 & 0.058 & 0.167 & 0.51 \\
\hline
\end{tabular}
\end{center}
%\vspace{-2ex}
\caption{Contributions of different perturbative orders to the partial and total $Z$ widths. A fixed value of $\MW$ has been used as input, 
instead of $G_\mu$. The $N_f$ and $N_f^2$ refer to corrections with 
one and two closed fermion loops, respectively, whereas $\alpha^2_{\rm bos}$
denotes contributions without closed fermion loops. Furthermore, $\alpha_\Pt$ 
and $\as$ are scale-dependent strong couplings.
% = y_\Pt^2/(4\pi)$.
%In all rows the radiator functions ${\cal R}_{\rm V,A}$ with known contributions
%through $\OO(\as^4)$, $\OO(\alpha^2)$ and $\OO(\alpha\as)$ are included.
%For input parameters and other details see 
Table from \cite{Dubovyk:2018rlg}.
\label{tab:res1}
}
\end{table*}
}}
\end{center}

Tab.~\ref{tab:res1} shows the results of higher order contributions to the Z-boson decay partial widths. Tab.~\ref{tab:err1} summarizes the estimation of the errors connected with unknown higher order corrections. For other EWPOs like $\seff{\ell}$,
$\seff{b}$, branching ratios, and the hadronic cross section at the Z-resonance, see \cite{Dubovyk:2016ocz,Dubovyk:2018rlg,Dubovyk:2019szj}. The total error for $\GZ$ in Tab.~\ref{tab:err1} amounts to 0.4 MeV, which is at the level of the CEPC accuracy ($~0.5$ MeV), while for the FCC-ee the experimental errors are estimated at the level of 0.1 MeV. That is why further progress in theoretical calculations is needed. In what follows we discuss recent developments in the numerical calculation of massive multi-loop Feynman integrals, in order to finally meet the future experimental demands. 
{\small{
% table 8, JHEP2019
\begin{table}[h!]
\centering
\renewcommand{\arraystretch}{1.3}
\begin{tabular}{|l|cccc|c|}
\hline
Observable & $\alpha\as^2$ & $\alpha\as^3$ & $\alpha^2\as$ & $\alpha^3$ & Total 
\\
\hline
$\Gamma_{e,\mu,\tau}$ [MeV] & 0.008 & 0.001 & 0.010 & 0.013 & 0.018 \\
$\Gamma_{\nu}$ [MeV] & 0.008 & 0.001 & 0.008 & 0.011 & 0.016 \\
$\Gamma_{u,c}$ [MeV] & 0.025 & 0.004 & 0.08 & 0.07 & 0.11 \\
$\Gamma_{d,s}$ [MeV] & 0.016 & 0.003 & 0.06 & 0.05 & 0.08 \\
$\Gamma_{b}$ [MeV] & 0.11 & 0.02 & 0.13 & 0.06 & 0.18 \\
$\GZ$ [MeV] & 0.23 & 0.035 & 0.21 & 0.20 & 0.4 \\
%\hline 
%$\seff{\ell}$ [$10^{-5}$] & --- & 0.3 & 3.0 & 3.1 & 4.3 \\
%$\seff{b}$ [$10^{-5}$] & 0.7 & 0.4 & 4.3 & 3.2 & 5.3 \\
\hline
\end{tabular}
\caption{Leading unknown higher-order corrections and their estimated order of magnitude for various pseudo-observables. The different orders always correspond to missing higher orders beyond the known approximations in the limit of a large top Yukawa coupling. The last column gives the total theory error obtained by adding the individual orders in quadrature. Table taken from  \cite{Dubovyk:2018rlg}.
\label{tab:err1}}
\end{table}
%-------------------------------------------------------------
%\section{State-of-the-art in the $Z-$boson decay multiloop numerical studies}
}}

 There are still no established general procedures for massive  {\em complete} perturbation theory calculations of Feynman integrals beyond one loop. For this reason, numerical integration methods are presently the most promising, if not the only, avenues
for addressing those challenges. Analytical techniques are expected to be important in many respects, but
numerical integration methods have advantages when increasing the number of masses and momentum
scales. Fortunately, there has been impressive progress in recent years in this direction \cite{Blondel:2018mad}. %There are
%currently two numerical methods known to allow a systematic treatment of infra-red divergences. 
In 2014 the only advanced automatic numerical two-loop method was sector decomposition (SD). However, the
corresponding software was not sufficiently developed to evaluate the complete set of Feynman integrals
for the massive electroweak bosonic two-loop corrections to the Z-boson decay with the desired high precision
(aiming at eight digits per integral). The task could be completed successfully with a substantial development of a
competing numerical approach, based on Mellin-Barnes (MB) representations of Feynman integrals \cite{Dubovyk:2016ocz}. These
calculations are challenging due to the numerical role of particle masses $M_Z$, $M_W$, $m_t$, $M_H$, leading to (i)
an enormous number of contributions, ranging from tens to hundreds of thousands of diagrams (at 3-loops),
and (ii) the occurrence of up to four dimensionless parameters in Minkowskian kinematics (at $s = M_Z^2$)
with intricate threshold and on-shell effects where contour deformation fails. In tackling more loops or legs,
merging both the MB- and SD-methods in numerical calculations, 
was the key for solving the complete massive SM two-loop case.
% is the key for ultimate success. 
We illustrate recent advances for multi-loop calculations applied to the Z-boson precision calculations using both methods. 

\begin{figure}[h!]
  \begin{center}
    \includegraphics[width=0.7\textwidth]{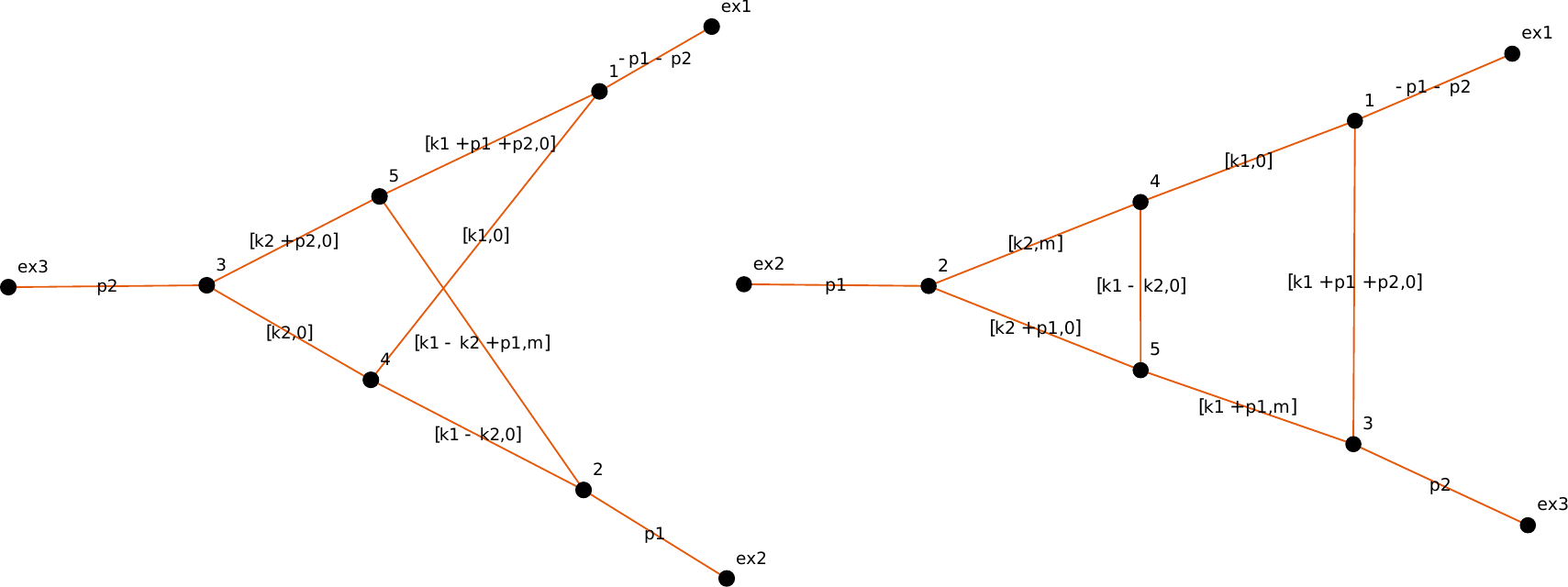}
 	\end{center}
  \caption{Left: Non-planar vertex with one massive crossed line. Right: Planar vertex with a finite part in the $\epsilon$ expansion represented by the single 3-dimensional \mbr{} integral of 
  Eqn.~(\ref{3dimEx11}).	 Figures generated by \pltest{} \cite{Bielas:2013rja,ambrewww}. Both vertices are special cases for which analytical solutions are available.  \label{ichepfig} 
}
\end{figure}

%The non-trivial diagrams which we will discuss are gathered in Fig.~\ref{ichepfig}. The \mb{} representation for the diagram on the left hand side is given in the file \verb+int15_MinkowskianMB.nb+ at  \cite{ambrewww}.
The non-trivial diagrams which we will discuss are gathered in 
Fig.~\ref{ichepfig}. The \mb{} representation for the non-planar diagram on the 
left hand side is four dimensional.
In this case, results obtained for the constant parts of the $\epsilon$-expansion with different methods and programs in the Euclidean region are, for  
$(p_1+p_2)^2=-m^2=-1$:

{
\begin{equation}
\begin{array}{ll}
{\text{ Analytical\; \cite{Fleischer:1998nb}:}} &       -{\color{red}{0.4966198306057021}} \\ 
{\text{ MB (Vegas)\; \cite{Czakon:2005rk}:}}   &      -{\color{red}{0.496}}9417442183914 \\     
{\text{ MB (Cuhre)\; \cite{Czakon:2005rk}:}}    &     -{\color{red}{0.49661983}}13219404 \\
{\text{ FIESTA\; \cite{Smirnov:2015mct}:}}     &       -{\color{red}{0.496618}}4488196595  \\ 
{\text{ SecDec\; \cite{Borowka:2017idc}:}}    &        -{\color{red}{0.49661983}}13167105 
\end{array} 
\end{equation}
}
\\
In the Minkowskian region, with  
$(p_1+p_2)^2=m^2=1$:
{\begin{equation}
\begin{array}{lc}
{\text{ Analytical \; \cite{Fleischer:1998nb}:}} &    -{\color{red}{0.778599608979684}} - {\color{red}{4.12351
2593396311}} \cdot i \\
{\text{ MBnumerics\;\cite{Dubovyk:2016aqv,Usovitsch:2018shx} }}:  &    -{\color{red}{0.778599608}}324769 - {\color{red}{4.123512}}600516016 \cdot i\\
%{\rm MB(QMC)}:     &    -{\color{red}{0.7785}}24251263640 - {\color{red}{4.12351
%2}}600516016 \cdot i\\
%{\rm MB(Vegas)}:   &                         {\rm big\; error} \\
{\rm MB (Cuhre)}:  &    -{\color{red}{0.7785}}24251263640 - {\color{red}{4.123}}498264231095 \cdot i\\
%{\rm FIESTA}:      &                         {\rm big\; error}\\
{\rm SecDec}:      &      {\rm big\; error\; {\bf [2016]}, {\color{red}{-0.77}} - i \cdot {\color{red}{4.1}}\;
 {\bf [2017]}, -{\color{red}{0.778}} - i \cdot {\color{red}{4.123}} \;{\bf [2019]}} \\%using splitting 
 {\rm pySecDec+rescaling  }:      & -{\color{red}{0.77859}}8 - i \cdot {\color{red}{4.123512}} \;{\bf [2020]}\\
\end{array}
\end{equation}
}
 The  SecDec group discussed this integral in \cite{Borowka:2017idc}. Using the splitting method the reported result is $-0.77 - i \cdot 4.1$. 
 For pySecDec with quasi-Monte Carlo integration (\texttt{QMC})
\cite{Borowka:2018goh} and using rescaling for $10^{7}$ generated points, the accuracy is much better.  Such integral is relatively easy for the \mb{} method, because it includes only one massive propagator.
The result for ${\rm MB (Cuhre)}$ has been obtained 
with the \mbm{} options: MaxPoints $10^7$, AccuracyGoal 8, PrecisionGoal 8. It took about 5 minutes on a moderate laptop. 

%in https://arxiv.org/pdf/1703.09692.pdf (p.16). They got
%0.77 + i*4.1 using splitting. Looking into python examples files
%one can find that they most likely used Divone with 10^7 points.
%So with the same amount of points, we have plus/minus the same accuracy, and the %advantage is only in speed.
% For more examples and comparisons, see \cite{tordll2016,Dubovyk:2016-tobepublished}.
Another interesting case is the planar scalar integral in Fig.~\ref{ichepfig}, right.

%\begin{figure}[h!]
	% \centering
%	\hspace*{3cm}
	%  \includegraphics[width=0.3\textwidth]{0h0w_b1}
	%  \includegraphics[width=0.3\textwidth]{0h0w_v1}
%	\includegraphics[width=0.4\textwidth]{figs/VertexMink.pdf}
%\caption{The integral with a finite part in $\epsilon$ expansion represented by a single 3-dim \mb{} integral (\ref{3dimEx11}).	 Figure generated by \pltest{} \cite{Bielas:2013rja,Bielas:2013v11}.\label{ichepfig}}
%\end{figure}
The MB representation for the constant term of this diagram is three-dimensional:
\begin{align}
I = \frac{1}{(2 \pi i)^3} \frac{1}{s^2}  
 \int\limits_{-i \infty - \frac{47}{37}}^{i \infty - \frac{47}{37}} dz_1 
 \int\limits_{-i \infty - \frac{44}{211}}^{i \infty - \frac{44}{211}} dz_2
 \int\limits_{-i \infty - \frac{176}{235}}^{i \infty - \frac{176}{235}} dz_3
 \left(\frac{m^2}{-s} \right)^{z_1}  \Gamma(-1 - z_1) \Gamma(2 + z_1)   \Gamma(-1 - z_{12})  \Gamma(-z_2)\nonumber \\
 \Gamma^2(1 + z_{12} - z_3) \Gamma(1 + z_3) 
  \Gamma(-z_3)  \Gamma^2(-z_1 + z_3) \Gamma(-z_{12} + z_3) 
   / \Gamma(-z_1) \Gamma(1 - z_2) \Gamma(1 - z_1 + z_3).
\label{3dimEx11} 
\end{align}
The diagram has also an analytical solution \cite{Aglietti:2004tq} which makes it ideal for a non-trivial comparison of different numerical techniques.
 	Numerical results for Eq.~\ref{3dimEx11} are presented in Tab.~\ref{tab:NumTabI2} for %kinematically the most demanded values 
 	$s = m^2 = 1$.
 	%\texttt{AS} stands for an analytical solution. 
 	
 \begin{table}[h!]
 	\centering
 	\begin{tabular}{|l|ll|l|} \hline
 		\rule{0pt}{2.3ex}\texttt{AS}      & $-1.{\bf199526183135}$ & $+5.{\bf567365907880} i$ &   \\ \hline 
 		%\rule{0pt}{2.3ex}\texttt{MB}$1$   & $-1.{\bf19952}5259137$ & %$+5.{\bf56736}7419371 i$ & Cuhre,  $10^7$, $10^{-8}$  \\ \hline
 		%\rule{0pt}{2.3ex}\texttt{MB}$2$   & $-1.{\bf19952}4318757$ & %$+5.{\bf567365}298565 i$ & Cuhre,  $10^7$, $10^{-8}$  \\ \hline
 		%\rule{0pt}{2.3ex}\texttt{MB}$3$   & $-1.{\bf199526}239547$ & %$+5.{\bf567365}843910 i$ & Cuhre,  $10^7$, $10^{-8}$  \\ \hline
 		\rule{0pt}{2.3ex}\texttt{MB}   & $-1.{\bf1995261831}68$ & $+5.{\bf567365907}904 i$ & Cuhre,  $10^7$, $10^{-8}$  \\ \hline
 	%	\rule{0pt}{2.3ex}\texttt{MB}$5$   & \texttt{NaN}           %&                          & Cuhre,  $10^7$             \\ \hline 
 		\rule{0pt}{2.3ex}\texttt{MB}   & $-1.{\bf20}4597845834$ & $+5.{\bf567}518701898 i$ & Vegas,  $10^7$, $10^{-3}$  \\ \hline
 %		\multicolumn{4}{|l|}{} \\ \hline
 		\rule{0pt}{2.3ex}\texttt{MB}   & $-1.{\bf1995}16455248$ & $+5.{\bf5673}76681167 i$ & QMC, $10^6$, $10^{-5}$     \\ \hline
 		\rule{0pt}{2.3ex}\texttt{MB}   & $-1.{\bf19952}7580305$ & $+5.{\bf56736}7345229 i$ & QMC, $10^7$, $10^{-6}$     \\ \hline
 	\end{tabular}
 		\caption{\label{tab:NumTabI2}
 		Numerical results for Eq.~\ref{3dimEx11} with  $s = m^2 = 1$. \texttt{AS} - analytical solution. For details on different  \texttt{MB}   integration routines and transformations of the infinite integration region used, see \cite{Dubovyk:2019krd}. Table taken from there, shortened.} 
 \end{table} 

Numerical results obtained for this integral have been discussed recently in \cite{Dubovyk:2019krd} with various transformations of variables and various deterministic and Monte Carlo integrators like the \texttt{CUHRE} routine, \texttt{VEGAS} routine \cite{Lepage:1977sw,Lepage:1980dq}, \texttt{QMC}. The \texttt{QMC} 
quasi-MC or \texttt{VEGAS} Monte Carlo methods surpass \texttt{CUHRE} for higher dimensional integrals. 
The \texttt{QMC} library seems to be especially suitable for the numerical integration of MB integrals in the Minkowskian region. It will be tested in more detail at the 3-loop level. The new \texttt{Vegas+} package \cite{Lepage:2020tgj} will be also studied.

In summary, there is substantial  progress in the numerical treatment of multi-loop Feynman integral calculations with  \mb{} and \texttt{SecDec}, approaching now the massive 3-loop diagrams. 
The techniques presented here can be extended for the computation of massive three-loop electroweak Feynman integrals needed for Z-peak physics.
It is also worth mentioning that the differential equations method \cite{Dlapa:2020cwj,Hidding:2020ytt} and the quoted \texttt{IBP} reductions are rapidly developing  \cite{Prausa:2020psw,Klappert:2020nbg}. They are expected to be very helpful, if not decisive for solving complete sets of integrals, as the third numerical method in the forthcoming three-loop studies.
%We are confident to finish in the foreseeable future all the necessary massive three-loop electroweak Feynman integrals needed for the $Z$-peak calculations. Saying this with other words: We see no showstoppers for this specific technical task.
%A short term ago, such a task seemed to be far beyond the reach. 
 Based on initial work in this direction we see no showstoppers for this specific technical task, and even though much additional work will be needed to assemble them into phenomenological results, this goal also appears within reach in the foreseeable future.

{\bf{Acknowledgments.}}

The work of \textit{A.F.}\  is supported in part by the National Science Foundation under
grant PHY-1820760. 
\textit{J.U.}\ received funding from the European Research Council (ERC) under the European Union's Horizon 2020 research and innovation programme under grant agreement no.\ 647356 (CutLoops).
The work is also supported in part by the Polish National Science Centre under
grant no.\ 2017/25/B/ST2/01987 and COST Action CA16201 PARTICLEFACE.


\begin{thebibliography}{10}
\expandafter\ifx\csname url\endcsname\relax
  \def\url#1{\texttt{#1}}\fi
\expandafter\ifx\csname urlprefix\endcsname\relax\def\urlprefix{URL }\fi
\expandafter\ifx\csname href\endcsname\relax
  \def\href#1#2{#2} \def\path#1{#1}\fi

\begin{minipage}{0.95\textwidth} \vskip .4\baselineskip
\bibitem{ALEPH:2005ab}
{ALEPH collab., DELPHI collab., L3 collab., OPAL collab., SLD collab., LEP
  Electroweak Working Group, SLD Electroweak Group, SLD Heavy Flavour Group, S.
  Schael}, et~al., {Precision electroweak measurements on the $Z$ resonance},
  Phys. Rept. 427 (2006) 257--454.
\newblock \href {http://arxiv.org/abs/hep-ex/0509008}
  {\path{arXiv:hep-ex/0509008}}, \href
  {http://dx.doi.org/10.1016/j.physrep.2005.12.006}
  {\path{doi:10.1016/j.physrep.2005.12.006}}.
\end{minipage}

\begin{minipage}{0.95\textwidth} \vskip .4\baselineskip
\bibitem{Bardin:1995-YR03}
{D. Bardin, W. Hollik, G. Passarino (eds.)}, {Reports of the working group on
  precision calculations for the $Z$ resonance, Yellow Report CERN 95-03
  (1995), parts I to III, 410 p.,
  \url{http://cds.cern.ch/record/280836/files/CERN-95-03.pdf}}.
\newblock \href {http://dx.doi.org/10.5170/CERN-1995-003}
  {\path{doi:10.5170/CERN-1995-003}}.
\end{minipage}

\begin{minipage}{0.95\textwidth} \vskip .4\baselineskip
\bibitem{Abada:2019zxq}
A.~Abada, et~al., {FCC-ee: The Lepton Collider}, Eur. Phys. J. ST 228~(2)
  (2019) 261--623.
\newblock \href {http://dx.doi.org/10.1140/epjst/e2019-900045-4}
  {\path{doi:10.1140/epjst/e2019-900045-4}}.
\end{minipage}

\begin{minipage}{0.95\textwidth} \vskip .4\baselineskip
\bibitem{Blondel:2019qlh}
A.~Blondel, A.~Freitas, J.~Gluza, T.~Riemann, S.~Heinemeyer, S.~Jadach,
  P.~Janot, {Theory Requirements and Possibilities for the FCC-ee and other
  Future High Energy and Precision Frontier Lepton Colliders, }\href
  {http://arxiv.org/abs/1901.02648} {\path{arXiv:1901.02648}}.
\end{minipage}

\begin{minipage}{0.95\textwidth} \vskip .4\baselineskip
\bibitem{CEPC-SPPCStudyGroup:2015csa}
M.~Ahmad, et~al., {CEPC-SPPC Preliminary Conceptual Design Report; Physics and
  Detector,
  }\href{http://inspirehep.net/record/1395734/files/main_preCDR.pdf}{http://inspirehep.net/record/1395734/files/main\_preCDR.pdf}.
\end{minipage}

\begin{minipage}{0.95\textwidth} \vskip .4\baselineskip
\bibitem{Blondel:2018mad}
A.~Blondel, et~al., {Standard model theory for the FCC-ee Tera-Z stage}, in:
  {Mini Workshop on Precision EW and QCD Calculations for the FCC Studies :
  Methods and Techniques}, Vol. 3/2019 of CERN Yellow Reports: Monographs,
  CERN, Geneva, 2018.
\newblock \href {http://arxiv.org/abs/1809.01830} {\path{arXiv:1809.01830}},
  \href {http://dx.doi.org/10.23731/CYRM-2019-003}
  {\path{doi:10.23731/CYRM-2019-003}}.
\end{minipage}

\begin{minipage}{0.95\textwidth} \vskip .4\baselineskip
\bibitem{Dubovyk:2016aqv}
I.~Dubovyk, A.~Freitas, J.~Gluza, T.~Riemann, J.~Usovitsch, {The two-loop
  electroweak bosonic corrections to $\sin^2\theta_{\rm eff}^{\rm b}$}, Phys.
  Lett. B762 (2016) 184--189.
\newblock \href {http://arxiv.org/abs/1607.08375} {\path{arXiv:1607.08375}},
  \href {http://dx.doi.org/10.1016/j.physletb.2016.09.012}
  {\path{doi:10.1016/j.physletb.2016.09.012}}.
\end{minipage}

\begin{minipage}{0.95\textwidth} \vskip .4\baselineskip
\bibitem{Dubovyk:2018rlg}
I.~Dubovyk, A.~Freitas, J.~Gluza, T.~Riemann, J.~Usovitsch, {Complete
  electroweak two-loop corrections to $Z$ boson production and decay}, Phys.
  Lett. B783 (2018) 86--94.
\newblock \href {http://arxiv.org/abs/1804.10236} {\path{arXiv:1804.10236}},
  \href {http://dx.doi.org/10.1016/j.physletb.2018.06.037}
  {\path{doi:10.1016/j.physletb.2018.06.037}}.
\end{minipage}

\begin{minipage}{0.95\textwidth} \vskip .4\baselineskip
\bibitem{Dubovyk:2019szj}
I.~Dubovyk, A.~Freitas, J.~Gluza, T.~Riemann, J.~Usovitsch, {Electroweak
  pseudo-observables and Z-boson form factors at two-loop accuracy}, JHEP 08
  (2019) 113.
\newblock \href {http://arxiv.org/abs/1906.08815} {\path{arXiv:1906.08815}},
  \href {http://dx.doi.org/10.1007/JHEP08(2019)113}
  {\path{doi:10.1007/JHEP08(2019)113}}.
\end{minipage}

\begin{minipage}{0.95\textwidth} \vskip .4\baselineskip
\bibitem{Dubovyk:2016ocz}
I.~Dubovyk, J.~Gluza, T.~Riemann, J.~Usovitsch, {Numerical integration of
  massive two-loop Mellin-Barnes integrals in Minkowskian regions}, PoS LL2016
  (2016) 034. \url{https://pos.sissa.it/260/034/pdf}.
\newblock \href {http://arxiv.org/abs/1607.07538} {\path{arXiv:1607.07538}},
  \href {http://dx.doi.org/10.22323/1.260.0034}
  {\path{doi:10.22323/1.260.0034}}.
\end{minipage}

\begin{minipage}{0.95\textwidth} \vskip .4\baselineskip
\bibitem{Bielas:2013rja}
K.~Bielas, I.~Dubovyk, J.~Gluza, T.~Riemann, {Some Remarks on Non-planar
  Feynman Diagrams}, Acta Phys. Polon. B44~(11) (2013) 2249--2255.
\newblock \href {http://arxiv.org/abs/1312.5603} {\path{arXiv:1312.5603}},
  \href {http://dx.doi.org/10.5506/APhysPolB.44.2249}
  {\path{doi:10.5506/APhysPolB.44.2249}}.
\end{minipage}

\begin{minipage}{0.95\textwidth} \vskip .4\baselineskip
\bibitem{ambrewww}
\texttt{AMBRE}~webpage: \url{http://prac.us.edu.pl/~gluza/ambre},\\
  Backup:~\href{https://web.archive.org/web/20200514010912/http://prac.us.edu.pl/~gluza/ambre/}{https://web.archive.org/web/20200514010912/http://prac.us.edu.pl/~gluza/ambre/}.
\end{minipage}

\begin{minipage}{0.95\textwidth} \vskip .4\baselineskip
\bibitem{Fleischer:1998nb}
J.~Fleischer, A.~Kotikov, O.~Veretin, {Analytic two loop results for selfenergy
  type and vertex type diagrams with one nonzero mass}, Nucl. Phys. B547 (1999)
  343--374.
\newblock \href {http://arxiv.org/abs/hep-ph/9808242}
  {\path{arXiv:hep-ph/9808242}}, \href
  {http://dx.doi.org/10.1016/S0550-3213(99)00078-4}
  {\path{doi:10.1016/S0550-3213(99)00078-4}}.
\end{minipage}

\begin{minipage}{0.95\textwidth} \vskip .4\baselineskip
\bibitem{Czakon:2005rk}
M.~Czakon, {Automatized analytic continuation of Mellin-Barnes integrals},
  Comput. Phys. Commun. 175 (2006) 559--571, mathematica program MB.m version
  1.2 (Jan 2, 2009), available at the MB Tools webpage,
  \url{http://projects.hepforge.org/mbtools/}.
\newblock \href {http://arxiv.org/abs/hep-ph/0511200}
  {\path{arXiv:hep-ph/0511200}}, \href
  {http://dx.doi.org/10.1016/j.cpc.2006.07.002}
  {\path{doi:10.1016/j.cpc.2006.07.002}}.
\end{minipage}

\begin{minipage}{0.95\textwidth} \vskip .4\baselineskip
\bibitem{Smirnov:2015mct}
A.~V. Smirnov, {FIESTA 4: Optimized Feynman integral calculations with GPU
  support}, Comput. Phys. Commun. 204 (2016) 189--199.
\newblock \href {http://arxiv.org/abs/1511.03614} {\path{arXiv:1511.03614}},
  \href {http://dx.doi.org/10.1016/j.cpc.2016.03.013}
  {\path{doi:10.1016/j.cpc.2016.03.013}}.
\end{minipage}

\begin{minipage}{0.95\textwidth} \vskip .4\baselineskip
\bibitem{Borowka:2017idc}
S.~Borowka, G.~Heinrich, S.~Jahn, S.~P. Jones, M.~Kerner, J.~Schlenk, T.~Zirke,
  {pySecDec: a toolbox for the numerical evaluation of multi-scale integrals},
  Comput. Phys. Commun. 222 (2018) 313--326.
\newblock \href {http://arxiv.org/abs/1703.09692} {\path{arXiv:1703.09692}},
  \href {http://dx.doi.org/10.1016/j.cpc.2017.09.015}
  {\path{doi:10.1016/j.cpc.2017.09.015}}.
\end{minipage}

\begin{minipage}{0.95\textwidth} \vskip .4\baselineskip
\bibitem{Usovitsch:2018shx}
J.~Usovitsch, I.~Dubovyk, T.~Riemann, {MBnumerics: Numerical integration of
  Mellin-Barnes integrals in physical regions}, PoS LL2018 (2018) 046.
\newblock \href {http://arxiv.org/abs/1810.04580} {\path{arXiv:1810.04580}},
  \href {http://dx.doi.org/10.22323/1.303.0046}
  {\path{doi:10.22323/1.303.0046}}.
\end{minipage}

\begin{minipage}{0.95\textwidth} \vskip .4\baselineskip
\bibitem{Borowka:2018goh}
S.~Borowka, G.~Heinrich, S.~Jahn, S.~P. Jones, M.~Kerner, J.~Schlenk, {A GPU
  compatible quasi-Monte Carlo integrator interfaced to pySecDec}, Comp. Phys.
  Comm., online.~\href {http://arxiv.org/abs/1811.11720}
  {\path{arXiv:1811.11720}}, \href
  {http://dx.doi.org/10.1016/j.cpc.2019.02.015}
  {\path{doi:10.1016/j.cpc.2019.02.015}}.
\end{minipage}

\begin{minipage}{0.95\textwidth} \vskip .4\baselineskip
\bibitem{Aglietti:2004tq}
U.~Aglietti, R.~Bonciani, {Master integrals with 2 and 3 massive propagators
  for the 2 loop electroweak form-factor - planar case}, Nucl. Phys. B698
  (2004) 277--318.
\newblock \href {http://arxiv.org/abs/hep-ph/0401193}
  {\path{arXiv:hep-ph/0401193}}, \href
  {http://dx.doi.org/10.1016/j.nuclphysb.2004.07.018}
  {\path{doi:10.1016/j.nuclphysb.2004.07.018}}.
\end{minipage}

\begin{minipage}{0.95\textwidth} \vskip .4\baselineskip
\bibitem{Dubovyk:2019krd}
I.~Dubovyk, J.~Gluza, T.~Riemann, {Optimizing the Mellin-Barnes Approach to
  Numerical Multiloop Calculations}, Acta Phys. Polon. B 50 (2019) 1993--2000.
\newblock \href {http://arxiv.org/abs/1912.11326} {\path{arXiv:1912.11326}},
  \href {http://dx.doi.org/10.5506/APhysPolB.50.1993}
  {\path{doi:10.5506/APhysPolB.50.1993}}.
\end{minipage}

\begin{minipage}{0.95\textwidth} \vskip .4\baselineskip
\bibitem{Lepage:1977sw}
G.~P. Lepage, {A New Algorithm for Adaptive Multidimensional Integration}, J.
  Comput. Phys. 27 (1978) 192.
\newblock \href {http://dx.doi.org/10.1016/0021-9991(78)90004-9}
  {\path{doi:10.1016/0021-9991(78)90004-9}}.
\end{minipage}

\begin{minipage}{0.95\textwidth} \vskip .4\baselineskip
\bibitem{Lepage:1980dq}
G.~P. Lepage, {VEGAS}: An adaptive multidimensional integration program,{
  }{\url{https://lib-extopc.kek.jp/preprints/PDF/1980/8006/8006210.pdf}}.
\end{minipage}

\begin{minipage}{0.95\textwidth} \vskip .4\baselineskip
\bibitem{Lepage:2020tgj}
G.~P. Lepage, {Adaptive Multidimensional Integration: VEGAS Enhanced, }\href
  {http://arxiv.org/abs/2009.05112} {\path{arXiv:2009.05112}}.
\end{minipage}

\begin{minipage}{0.95\textwidth} \vskip .4\baselineskip
\bibitem{Dlapa:2020cwj}
C.~Dlapa, J.~Henn, K.~Yan, {Deriving canonical differential equations for
  Feynman integrals from a single uniform weight integral}, JHEP 05 (2020) 025.
\newblock \href {http://arxiv.org/abs/2002.02340} {\path{arXiv:2002.02340}},
  \href {http://dx.doi.org/10.1007/JHEP05(2020)025}
  {\path{doi:10.1007/JHEP05(2020)025}}.
\end{minipage}

\begin{minipage}{0.95\textwidth} \vskip .4\baselineskip
\bibitem{Hidding:2020ytt}
M.~Hidding, {DiffExp, a Mathematica package for computing Feynman integrals in
  terms of one-dimensional series expansions, }\href
  {http://arxiv.org/abs/2006.05510} {\path{arXiv:2006.05510}}.
\end{minipage}

\begin{minipage}{0.95\textwidth} \vskip .4\baselineskip
\bibitem{Prausa:2020psw}
M.~Prausa, J.~Usovitsch, {The analytic leading color contribution to the
  Higgs-gluon form factor in QCD at NNLO, }\href
  {http://arxiv.org/abs/2008.11641} {\path{arXiv:2008.11641}}.
\end{minipage}

\begin{minipage}{0.95\textwidth} \vskip .4\baselineskip
\bibitem{Klappert:2020nbg}
J.~Klappert, F.~Lange, P.~Maierh\"ofer, J.~Usovitsch, {Integral Reduction with
  Kira 2.0 and Finite Field Methods, }\href {http://arxiv.org/abs/2008.06494}
  {\path{arXiv:2008.06494}}.
\end{minipage}

\end{thebibliography}
\end{document}